\documentclass[prd,aps,floats,twocolumn]{revtex4}
\newcommand{\be}{\begin{equation}}
\newcommand{\ee}{\end{equation}}
\newcommand{\ba}{\begin{eqnarray}}
\newcommand{\ea}{\end{eqnarray}}
\newcommand{\bc}{}

\begin{document}

\preprint{\
\begin{tabular}{rr}
&
\end{tabular}
}
\title{The Vector-Tensor nature of Bekenstein's relativistic theory 
of Modified Gravity }
\author{T.G.~Zlosnik$^{1}$, P.G.~Ferreira$^{2,3}$, and Glenn D.~Starkman
$^{4}$}
%
\affiliation{
$^1$Imperial College Theoretical Physics, Huxley Building, London SW7\\
$^{2}$Astrophysics, University of Oxford, Denys Wilkinson Building, Keble 
Road, Oxford OX1 3RH, UK\\
$^{3}$African Institute for Mathematical Sciences, 6-8 Melrose Road, Muizenberg 7945, South Africa\\
$^{4}$Department of Astronomy and Physics, Case Western Reserve, Cleveland, Ohio, U.S.A.}

\begin{abstract}
Bekenstein's theory of relativistic gravity is conventionally  written as a
bi-metric theory. The two metrics are related by a disformal transformation
defined by a dynamical vector field and a scalar field. In this comment we 
show that the theory can be re-written as Vector-Tensor theory akin to
Einstein-Aether theories with non-canonical kinetic terms. We discuss
some of the implications of this equivalence.
\end{abstract}

\date{\today}
\pacs{PACS Numbers : }
\maketitle

\noindent

There is a distinct possibility that Newtonian gravity, and its relativistic
generalization, General Relativity, breaks down in regions of small Newtonian
acceleration. Modified Newtonian Dynamics (MOND) proposes a fix to
the law of gravity in the non-relativistic regime which fares well
in explaining the dynamics of galaxies. In the past few years,
a relativistic generalization of MOND has been proposed by Bekenstein
with interesting consequences on cosmological scales. 

One might expect that a modified theory of gravity must tamper with
the way the gravitational field (or metric), couples or responds to
sources. Modifications of gravity typically involve modifying
the Einstein-Hilbert action or introducing extra degrees of 
freedom (or fields) that distort the way the metric enters the
action for all forms of matter. A well known example is Jordan-Brans-Dicke
theory where an extra scalar field can be interpreted as a time-varying
Newton's ``constant''. Such a theory can be rewritten (or transformed) with a
redefinition of the metric in such a way that Newton's ``constant'' becomes
constant but the matter action picks up couplings to the scalar field.

 More generally one can think of such theories as having two metrics. 
One metric
satisfies the Einstein-Hilbert action while the other defines the
stress-energy tensory and the geodesic equations. A rule must then
be posited that links the two metrics which typically involves
new dynamical fields with their own actions. Bekenstein's theory
falls in this class of theories.

Bekenstein's theory can be described as follows. Gravity is mediated 
by three fields: a tensor field $\tilde{g}_{ab}$ with associated 
metric-compatible connection $\tilde{\nabla}_{a}$ and well defined 
inverse $\tilde{g}^{ab}$, a timelike one-form field $A_{a}$ such 
that $\tilde{g}^{ab}A_{a}A_{b}=-1$, and a scalar field $\phi$. 
As advertised above, the metric $\tilde{g}_{ab}$ has its dynamics governed by 
the Einstein-Hilbert action,
\begin{eqnarray}
S_{\tilde{g}}= \frac{1}{16\pi G}\int d^{4}
x(-\tilde{g})^{\frac{1}{2}}\tilde{R} \nonumber
\end{eqnarray} 
where $G$ is Newton's constant and $\tilde{R}$ is the scalar curvature 
of $\tilde{g}_{ab}$. We shall call the frame of this metric 
the ``Einstein Frame'' (EF). The scalar $\phi$ has its dynamics given by
\begin{eqnarray}
S_{s} = -\frac{1}{16\pi G}\int d^{4}x(-\tilde{g})^{\frac{1}{2}}
\left[\mu(\tilde{g}^{ab}-A^{a}A^{b})\tilde{\nabla}_{a}
\phi\tilde{\nabla}_{b}\phi+V(\mu)\right] \nonumber
\end{eqnarray}
where we have used the convention employed in \cite{SKORD}.
$\mu$ is a non-dynamical field and $V$ is a free function which 
may be chosen so as to give the correct non-relativistic 
MONDian limit. The one-form field, $A_{a}$ has dynamics given by
\begin{eqnarray}
S_{v} = -\frac{1}{32\pi G}\int d^{4}
x(-\tilde{g})^{\frac{1}{2}}\left[K_{B}F^{ab}F_{ab}
-2\lambda(\tilde{g}^{ab}A_{a}A_{b}+1)\right] \nonumber
\end{eqnarray}
where $F_{ab}= 2\tilde{\nabla}_{[a}A_{b]}$ and brackets denote 
antisymmetrization. Indices are raised with $\tilde{g}_{ab}$ and 
$K_{B}$ is a dimensionless parameter. Variation with respect to 
the Lagrange multiplier field $\lambda$ yields the `unit timelike' constraint on $A_{a}$:
\begin{equation}
\label{constraint}
\tilde{g}^{ab}A_{a}A_{b}=-1 \nonumber
\end{equation}

Matter is required to obey the weak equivalence principle, 
which means that there is a metric $g_{ab}$ with associated 
metric-compatible connection $\nabla_{a}$, universal to all 
matter fields, such that test particles follow its geodesics. 
We shall call the frame of this metric the ``Matter Frame'' (MF). 
For some collection of matter fields $f^{A}$ the action is thus
\begin{eqnarray}
S_{m} = \int d^{4}x(-g)^{\frac{1}{2}}L\left[g,f^{A},
\partial f^{A}\right] \nonumber
\end{eqnarray}

When the unit-timelike constraint on $A_{a}$ is satisfied,
the metrics, one-form field, and scalar field are related by:
\begin{equation}
\label{md}
\tilde{g}_{ab} = e^{2\phi}g_{ab}+2e^{2\phi}A_{a}A_{b}\sinh(2\phi)
\end{equation}
and 
\begin{equation}
\label{mu}
\tilde{g}^{ab} =e^{-2\phi}g^{ab}-2e^{2\phi}g^{ac}g^{bd}A_{c}A_{d}\sinh(2\phi)
\end{equation}
The use of Lagrange's method of undetermined multipliers to realize 
the constraint (\ref{constraint}) in TeVeS means that (\ref{md}) 
and (\ref{mu}) do not apply at the level of the action
\footnote[1]{Note that we use action to refer to the integrals 
presented and not specifically to their form when the equations of motion are satisfied.}. 
In this form one may not write the action entirely in the MF or EF. 
However, in a variational problem with an algebraic constraint 
one may instead realize the constraint in the action itself and 
eliminate one of the variables \cite{lancz}. In doing so, 
(\ref{md}) and (\ref{mu}) may then be used in the action.

With a view to writing the action entirely in the MF, we may 
express the unit-timelike constraint in terms of 
$g^{ab}$, $A_{a}$, and $\phi$:
\begin{equation}
\label{mfcon}
A^{2} \equiv g^{ab}A_{a}A_{b} = -e^{-2\phi}
\end{equation}
This enables us to eliminate one degree of freedom: 
a simple choice is to eliminate the field $\phi$. The relations between the metrics now take the form
\begin{equation}
\label{md2}
\tilde{g}_{ab} = -\frac{1}{A^{2}}g_{ab}-\frac{1}{A^{2}}A_{a}A_{b}(A^{2}-\frac{1}{A^{2}})
\end{equation}
and 
\begin{equation}
\label{mu2}
\tilde{g}^{ab} =-A^{2}g^{ab}+\frac{1}{A^{2}}g^{ac}g^{bd}A_{c}A_{d}(A^{2}-\frac{1}{A^{2}})
\end{equation}
Knowing the relation between $g_{ab}$ and $\tilde{g}_{ab}$, we may 
relate their determinants and connections. As shown in \cite{BEK}, 
the determinants $g$ and $\tilde{g}$ are related as:
\begin{eqnarray}
\label{det}
(-\tilde{g})^{\frac{1}{2}}= e^{2\phi}(-g)^{\frac{1}{2}}=
-\frac{1}{A^{2}}(-g)^{\frac{1}{2}} 
\end{eqnarray}
where again we have used the constraint (\ref{mfcon}).

It may be shown \cite{WALD} that for any two derivative 
operators $\tilde{\nabla}_{a}$ and $\nabla_{a}$ there exists 
a tensor field $C^{c}_{\phantom{c}ab}$ such that for any one-form $\omega_{a} \in T^{*}M$
\begin{eqnarray}
\label{trans}
\tilde{\nabla}_{a}\omega_{b} =\nabla_{a}\omega_{b}-
C^{c}_{\phantom{c}ab}\omega_{c} \nonumber
\end{eqnarray}
where
\begin{eqnarray}
C^{c}_{\phantom{c}ab} =\frac{1}{2}\tilde{g}^{cd}
\left[\nabla_{a}\tilde{g}_{bd}+\nabla_{b}\tilde{g}_{ad}-
\nabla_{d}\tilde{g}_{ab}\right] \nonumber
\end{eqnarray}
and $TM$ ($T^{*}M$) are the tangent (co-tangent) space of the manifold $M$.

Beginning from the definition of the Riemann tensor 
$\tilde{R}_{abd}^{\phantom{abd}c}$ (for some vector $V^{a}\in TM$)
\begin{eqnarray}
(\tilde{\nabla}_{a}\tilde{\nabla}_{b}-
\tilde{\nabla}_{b}\tilde{\nabla}_{a})V^{c} = 
-\tilde{R}_{abd}^{\phantom{abd}c}V^{d} \nonumber
\end{eqnarray}
we may use (\ref{trans}) to write this in terms of 
$\nabla_{a}$ and $C^{a}_{\phantom{a}bd}$:
\begin{eqnarray}
\tilde{R}_{abc}^{\phantom{abc}d} = R_{abc}^{\phantom{abc}d} -
2\nabla_{[a}C^{d}_{\phantom{d}b]c}+
2C^{e}_{\phantom{e}c[a}C^{d}_{\phantom{d}b]e} \nonumber
\end{eqnarray}
and so we have that
\begin{eqnarray}
\label{scacur}
\tilde{R}= \tilde{g}^{ac}\left[R_{aec}^{\phantom{abc}e} -2\nabla_{[a}C^{e}_{\phantom{d}e]c}+
2C^{e}_{\phantom{e}c[a}C^{d}_{\phantom{d}d]e}\right] 
\end{eqnarray}
Therefore, with (\ref{det}) and (\ref{scacur}) we have all 
we need to rewrite $S_{\tilde{g}}$ in the matter frame. 
After some algebra we find that
\begin{eqnarray}
S_{\tilde{g}}= \frac{1}{16\pi G}
\int d^{4}x(-g)^{\frac{1}{2}}\left[R+K^{abmn}\nabla_{a}A_{m}
\nabla_{b}A_{n}\right] 
\end{eqnarray}
where
\begin{eqnarray}
K^{abmn} &=& d_{1}g^{ab}g^{mn} + d_{2}g^{am}g^{bn}+
d_{3}g^{an}g^{bm} \nonumber\\
\nonumber && +d_{4}A^{a}A^{b}g^{mn}+d_{5}
g^{an}A^{b}A^{m}+d_{6}g^{ab}A^{m}A^{n}\\
&& +d_{7}g^{am}A^{b}A^{n}+
d_{8}A^{a}A^{b}A^{m}A^{n}
\end{eqnarray}
and
\begin{eqnarray}
d_{1} &=& \left(-\frac{1}{2A^{6}}+\frac{1}{A^{2}}-\frac{1}{2}A^{2}\right)\\ 
d_{2} &=&\left(\frac{1}{A^{6}}-\frac{1}{A^{2}}\right) \\
d_{3} &=&-\frac{1}{2}\left(\frac{1}{A^{6}}-A^{2}\right)  \\
d_{4} &=& \left(\frac{1}{2A^{8}}-\frac{1}{A^{4}}+\frac{1}{2}\right) \\
d_{5} &=& \left(\frac{1}{A^{8}}+\frac{4}{A^{4}}-1\right)\\
d_{6} &=& \left(\frac{1}{2A^{8}}-\frac{3}{A^{4}}+\frac{1}{2}\right)  \\
d_{7} &=&\left(-\frac{6}{A^{8}}+\frac{2}{A^{4}}\right) \\
d_{8} &=&\left(\frac{10}{A^{10}}-\frac{2}{A^{6}}\right)
\end{eqnarray}
All indices have been raised with $g^{ab}$ and total 
divergence terms have been dropped. Rewriting the scalar action in the MF we find:
\begin{eqnarray}
S_{s} &=& -\frac{1}{16\pi G}\int d^{4}x(-g)^{\frac{1}{2}}
[\mu(e_{6}g^{ab}A^{m}A^{n}+\nonumber\\
&&
e_{8}A^{a}A^{b}A^{m}A^{n})\nabla_{a}A_{m}\nabla_{b}A_{n}-\frac{V(\mu)}{A^{2}}]
\end{eqnarray}
where 
\begin{eqnarray}
e_{6} &=& \frac{1}{A^{4}}  \\
e_{8} &=& \left(\frac{2}{A^{10}}-\frac{1}{A^{6}}\right)
\end{eqnarray}
Rewriting the vector action in the MF we find:

\begin{eqnarray}
S_{v} &=& \frac{K_{B}}{16\pi G}\int d^{4}x(-g)^{\frac{1}{2}}[
A^{2}(g^{ab}g^{mn}-g^{an}g^{mb})\nonumber \\
&& +\left(\frac{1}{A^{4}}-1\right)(g^{ab}A^{m}A^{n}
+g^{mn}A^{a}A^{b}-2g^{an}A^{m}A^{b})]\nonumber \\
&&\nabla_{a}A_{m}\nabla_{b}A_{n}
\end{eqnarray}
Therefore the total TeVeS action written entirely in the matter frame is:

\begin{eqnarray}
S_{T} &=& \frac{1}{16\pi G}\int d^{4}x(-g)^{\frac{1}{2}}
          \left[R+\check{K}^{abmn}\nabla_{a}A_{m}\nabla_{b}A_{n}
          +\frac{V(\mu)}{A^{2}}\right]\nonumber \\
          &&+S_{m}[g^{ab}]
\end{eqnarray}
where $\check{K}^{abmn}$ is of the same form as $K^{abmn}$ but with coefficients $\check{d}_{i}$ given by:

\begin{eqnarray}
\check{d}_{1}&=& d_{1}+K_{B}A^{2};\quad \check{d}_{2}=d_{2};\quad \check{d}_{3}=d_{3}-K_{B}A^{2} \nonumber \\
\check{d}_{4} &=& d_{4}+K_{B}\left(\frac{1}{A^{4}}-1\right);\quad \check{d}_{5}=d_{5}+2K_{B}\left(1-\frac{1}{A^{4}}\right) \nonumber \\
\check{d}_{6} &=&d_{6}+\frac{K_{B}-\mu}{A^{4}}-K_{B};\quad \check{d}_{7}=d_{7}; \nonumber \\
\check{d}_{8}&=&d_{8}-\frac{2\mu}{A^{10}}+\frac{\mu}{A^{6}}
\end{eqnarray}
We have thus succeeded in rewriting the total action as a functional of a 
single metric $g^{ab}$, the one form field $A_{a}$, 
the non-dynamical field $\mu$ and the matter fields. 
We now check that this action indeed produces the same
nonrelativistic limit as deduced in \cite{BEK}. Firstly we must obtain the field equations in the 
matter frame. Varying $\mu$ we find:

\begin{equation}
\left[\frac{1}{A^{2}}g^{ab}A^{m}A^{n}+(\frac{2}{A^{8}}-\frac{1}{A^{4}})
      A^{a}A^{b}A^{m}A^{n}\right]\nabla_{a}A_{m}\nabla_{b}A_{d}=\frac{dV}{d\mu}
\end{equation}
Varying $A_{a}$ we find:

\begin{equation}
\nabla_{a}\left[\check{K}^{aebc}\nabla_{b}A_{c}\right]
         =\frac{1}{2}J^{abcde}\nabla_{a}A_{b}\nabla_{c}A_{d}-\frac{1}{A^{4}}V(\mu)A^{e}
\end{equation}
where 

\begin{eqnarray*}
J^{abmne} \equiv \frac{\delta \check{K}^{abmn}}{A_{e}}
\end{eqnarray*}
Finally, varying $g^{ab}$ the Einstein equations are:

\begin{eqnarray}
G_{ab} &=& \frac{1}{2}\check{K}^{efcd}\nabla_{e}A_{f}\nabla_{c}A_{d} g_{ab}
           -S^{efcd}_{\phantom{efcd}ab}\nabla_{e}A_{f}\nabla_{c}A_{d} \nonumber \\
\nonumber && +\left[A_{a}J^{e\phantom{b}cdf}_{\phantom{e}b}+A_{b}J_{a}^{\phantom{a}ecdf}
                   -A^{e}J_{ab}^{\phantom{ab}cdf}\right]\nabla_{e}A_{f}\nabla_{c}A_{d} \nonumber \\
\nonumber && +\left[A_{a}\check{K}^{e\phantom{b}cd}_{\phantom{e}b}
                    +A_{b}\check{K}_{a}^{\phantom{a}ecd}
                    -A^{e}\check{K}_{ab}^{\phantom{ab}cd}\right]\nabla_{e}\nabla_{c}A_{d} \nonumber \\
\nonumber && +\left[\check{K}^{e\phantom{b}cd}_{\phantom{e}b}\nabla_{e}A_{a}
                    +\check{K}_{a}^{\phantom{a}ecd}\nabla_{e}A_{b}-\check{K}_{ab}^{\phantom{ab}cd}
                    \nabla_{e}A^{e}\right]\nabla_{c}A_{d} \nonumber \\
\nonumber && +\left[A_{a}M^{e\phantom{b}cd}_{\phantom{e}b}
                    +A_{b}M_{a}^{\phantom{a}ecd}
                    -A^{e}M_{ab}^{\phantom{ab}cd}\right]\nabla_{e}\mu \nabla_{c}A_{d} \nonumber
                     \\
 &&+\frac{V(\mu)}{A^{4}}A_{a}A_{b}+\frac{V(\mu)}{2A^{2}}g_{ab}+8\pi G T_{ab}
\end{eqnarray}
where

\begin{eqnarray*}
M^{abcd} &\equiv & -\frac{1}{A^{4}}g^{ac}A^{b}A^{d}+\left(\frac{1}{A^{6}}-\frac{2}{A^{10}}\right)
                 A^{a}A^{b}A^{c}A^{d}  \\
S^{abmn}_{\phantom{abmn}ef} &\equiv& \frac{\delta\check{K}^{abmn}}{\delta g^{ef}}
\end{eqnarray*}
We will now show that these field equations indeed yield TeVeS's MONDian behaviour in the non-relativistic regime. Towards these ends, consider a space-time admitting the following ansatz for the metric and one-form field: 

\begin{eqnarray}
g_{ab}&=&\eta_{ab}+\epsilon h_{ab}\\
A_{a} &=& -\delta^{0}_{\phantom{0}a}-\epsilon B_{a}
\end{eqnarray}
where $\eta_{ab}=\mathrm{diag}(-1,1,1,1)$ and $\epsilon$ keeps track of the order of perturbation. We will neglect  terms of order $\epsilon^{2}$ and above in the field equations. We will also neglect time 
derivatives of the two fields. We take $h_{00}=-2\Phi$ and $h_{ij}= -2\Phi\delta_{ij}$. This is an appropriate form of the metric perturbation as long as the vector field has zero curl to first order in $\epsilon$ and the presence of gravitational waves can be neglected. It may be checked that non-vanishing `$\nabla A \nabla A$' contributions to 
the field equations are at least of order two in $\epsilon$ 
and so can be neglected. Also, in this limit the coefficients 
$d_{i}(A^{2})$ and $e_{i}(A^{2})$ assume their values at $A^{2}=-1$. 

Varying the total action with respect to $g^{ab}$ we find that, up to 
$O(\epsilon)$, the time-time component of the Einstein equation is as follows:

\begin{eqnarray}
\nabla^{2}\Phi &=& 8\pi G \rho+(-2(\check{d}_{1}+\check{d}_{3})+(\check{d}_{4}+\check{d}_{5}
                            +\check{d}_{6}))\nabla^{2}\Phi \nonumber \\
&& -(-(\check{d}_{1}+\check{d}_{3})+\frac{1}{2}\check{d}_{5}+\check{d}_{6})\nabla^{2}B_{0}
\end{eqnarray}
It is shown in 
\cite{BEK} that one may safely regard contributory terms approximately 
equal to $V(\mu)$ as of order $\epsilon^{2}$.
Similarly, the vector equation is:

\begin{eqnarray}
0=(\check{d}_{1}+\check{d}_{3}-\frac{1}{2}\check{d}_{5}-\check{d}_{6})\nabla^{2}\Phi
  -(\check{d}_{1}-\check{d}_{6})\nabla^{2}B_{0}
\end{eqnarray}
In the limit of $A^{2}\rightarrow -1$, we  find
$\check{d}_{2}=\check{d}_{3}=0$, $\check{d}_{1}=-\check{d}_{3}=-K_{B}$, $\check{d
}_{5}=4$, $\check{d}_{6}=-(2+\mu)$. Substituting the vector equation 
into the time-time $g^{ab}$ equation we find:
\begin{equation}
\left(1-\frac{K_{B}}{2}\right)\nabla^{2}B_{0} = 4\pi G\rho 
\label{eq:linear00}
\end{equation}
This is the result obtained in \cite{BEK} i.e. that the EF `time-time' metric perturbation approximately obeys 
Poisson's equation up to a tracking component produced by a one-form field with fixed norm according to $\tilde{g}^{ab}A_{a}A_{b}=-1$. 
Therefore we may take $B_{0}=(1-\frac{K_{B}}{2})^{-1}\Phi_{N}$. 
Substituting this into the vector equation we find:
\begin{equation}
8\pi G\rho=\mu\nabla^{2}\left(\Phi-\frac{1}{1-\frac{K_{B}}{2}}\Phi_{N}\right)
\label{eq:poissonmod}
\end{equation}
It may be checked that the function $\Phi-\frac{1}{1-\frac{K_{B}}{2}}\Phi_{N}$ is equal to the scalar field $\phi$. This is indeed the scalar field equation found in \cite{BEK}. 
For a chosen $V$ one may invert this equation to obtain, schematically, $\mu=\mu(\nabla A \nabla A)$.
One may then find an appropriate $V$ so that (\ref{eq:poissonmod}) and (\ref{eq:linear00}) collectively produce a relationship between $\Phi$ and $\rho$ as given by the Bekenstein-Milgrom equation.

Of course one may equally have obtained the resulting behaviour in (\ref{eq:poissonmod}) by allowing for $S_{s}$ to be constructed from some function $F(W)$ of the kinetic terms $W$ contained therein. One may use either $\mu$ or F so long as an identification $\mu = \frac{dF}{dW}$ may be made.

The equivalence between Bekenstein's theory and the Vector-Tensor theory
we have presented above is intriguing. Due to the particular symmetries
of the transformation (and the time-like constraint on the vector field)
it is possible to show that this is not a true bimetric theory of
gravity. The same metric which satisfies the Einstein-Hilbert action
couples minimally to the matter fields. Yet it is still possible
to obtain modifications to gravity through the coupling of the metric
to the vector field. The Lorentz structure of the vector field 
is such that non-canonical kinetic terms will generate non-minimal
couplings to the Ricci and the Riemman tensors and hence to modified
field equations. 

Although the action written purely in terms of the vector and tensor
fields seems more intricate than the action originally proposed by Bekenstein,
it is conceptually simpler. Modifications to gravity can be seen to
arise through the sole existence of a dynamical ``aether'' field with 
particular properties. The simplest cases of Einstein-Aether theories 
of gravity have been extensively studied as have proposals for its origin, 
from fundamental physics to effective field theories \cite{JM,G}. 
Bekenstein's theory
can be seen as an extensions of such theories and hence amenable
to the same types of analysis.

The forms of the recovered equations suggests that a substantial simplification of TeVeS is possible in the
Einstein-Aether context.  We see that only one of $d_3$
and $d_5$ plus $e_6$ really need be non-zero, in order to arrive
at MOND phenomenology, and they might well be constants rather than
complicated functions of $A^2$. In addition, one needs to ensure that
$A^2=-1$ is the preferred solution around which to expand.
The latter might be enforced by retaining any of the other $d_i$
as appropriate functions of $A^2$, or by introducing an appropriate
potential for $A^2$.

{\it Acknowledgments}: We thank J.~Bekenstein and for useful 
comments and C.~Skordis for discovering a propagating error in the original draft. TGZ is funded by a PPARC studentship. GDS is funded by a 
Guggenheim fellowship and a Beecroft fellowship.


\end{document}